\documentclass[]{nature}
\makeatletter\if@twocolumn\PassOptionsToPackage{switch}{lineno}\else\fi\makeatother
\usepackage{gensymb}
\usepackage{tabulary,graphicx,amsmath,amsfonts,amssymb}
\usepackage[utf8x]{inputenc}

\makeatletter
\renewenvironment{figure}
               {\@float{figure}}
               {\end@float}
\renewenvironment{figure*}
               {\@dblfloat{figure}}
               {\end@dblfloat}

\renewenvironment{table*}
               {\@dblfloat{table}}
               {\end@dblfloat}

\makeatother

\usepackage{url,multirow,morefloats,floatflt,cancel,tfrupee}
\makeatletter

\AtBeginDocument{\@ifpackageloaded{textcomp}{}{\usepackage{textcomp}}}
\makeatother
\usepackage{colortbl}
\usepackage{xcolor}
\usepackage{pifont}
\usepackage[nointegrals]{wasysym}
\makeatletter

\def\mcWidth#1{\csname TY@F#1\endcsname+\tabcolsep}

\def\cAlignHack{\rightskip\@flushglue\leftskip\@flushglue\parindent\z@\parfillskip\z@skip}
\def\rAlignHack{\rightskip\z@skip\leftskip\@flushglue \parindent\z@\parfillskip\z@skip}

\@ifundefined{etal}{}{}

\usepackage{ifxetex}
\ifxetex\else\if@twocolumn\@ifpackageloaded{stfloats}{}{\usepackage{dblfloatfix}}\fi\fi

\AtBeginDocument{
\expandafter\ifx\csname eqalign\endcsname\relax
\def\eqalign#1{\null\vcenter{\def\\{\cr}\openup\jot\m@th
  \ialign{\strut$\displaystyle{##}$\hfil&$\displaystyle{{}##}$\hfil
      \crcr#1\crcr}}\,}
\fi
}

\AtBeginDocument{%
  \@ifpackageloaded{endfloat}%
   {\renewcommand\efloat@iwrite[1]{\immediate\expandafter\protected@write\csname efloat@post#1\endcsname{}}}{\newif\ifefloat@tables}%
}%

\def\BreakURLText#1{\@tfor\brk@tempa:=#1\do{\brk@tempa\hskip0pt}}
\let\lt=<
\let\gt=>
\def\processVert{\ifmmode|\else\textbar\fi}

\@ifundefined{subparagraph}{
\def\subparagraph{\@startsection{paragraph}{5}{2\parindent}{0ex plus 0.1ex minus 0.1ex}%
{0ex}{\normalfont\small\itshape}}%
}{}

\newcommand\role[1]{\unskip}
\newcommand\aucollab[1]{\unskip}

\@ifundefined{tsGraphicsScaleX}{\gdef\tsGraphicsScaleX{1}}{}
\@ifundefined{tsGraphicsScaleY}{\gdef\tsGraphicsScaleY{.9}}{}
\def\checkGraphicsWidth{\ifdim\Gin@nat@width>\linewidth
	\tsGraphicsScaleX\linewidth\else\Gin@nat@width\fi}

\def\checkGraphicsHeight{\ifdim\Gin@nat@height>.9\textheight
	\tsGraphicsScaleY\textheight\else\Gin@nat@height\fi}

\def\fixFloatSize#1{}
\let\ts@includegraphics\includegraphics

\def\inlinegraphic[#1]#2{{\edef\@tempa{#1}\edef\baseline@shift{\ifx\@tempa\@empty0\else#1\fi}\edef\tempZ{\the\numexpr(\numexpr(\baseline@shift*\f@size/100))}\protect\raisebox{\tempZ pt}{\ts@includegraphics{#2}}}}

\AtBeginDocument{\def\includegraphics{\@ifnextchar[{\ts@includegraphics}{\ts@includegraphics[width=\checkGraphicsWidth,height=\checkGraphicsHeight,keepaspectratio]}}}

\DeclareMathAlphabet{\mathpzc}{OT1}{pzc}{m}{it}

\def\URL#1#2{\@ifundefined{href}{#2}{\href{#1}{#2}}}

\def\UrlOrds{\do\*\do\-\do\~\do\'\do\"\do\-}%
\g@addto@macro{\UrlBreaks}{\UrlOrds}

\edef\fntEncoding{\f@encoding}

\makeatother

\newif\ifmultipleabstract\multipleabstractfalse%
\usepackage[nomarkers,tablesfirst,nolists]{endfloat}

\def\fixFloatSize#1{}

\title{Wide-field high-resolution 3D microscopy with Fourier ptychographic diffraction tomography}

\author{Chao Zuo$^{1,2*\dagger}$, Jiasong Sun$^{1,2\dagger}$, Jiaji Li$^{1,2}$, Anand Asundi$^3$, and Qian Chen$^{2*}$}

\begin{document}

\maketitle

\begin{affiliations}
 \item Smart Computational Imaging (SCI) Laboratory, Nanjing University of Science and Technology, Nanjing, Jiangsu Province 210094, China
 \item Jiangsu Key Laboratory of Spectral Imaging $\&$ Intelligent Sense, Nanjing University of Science and Technology, Nanjing, Jiangsu Province 210094, China
 \item Centre for Optical and Laser Engineering (COLE), School of Mechanical and Aerospace Engineering, Nanyang Technological University, Singapore 639798, Singapore

     $\dagger$Chao Zuo and Jiasong Sun contributed equally to this work.

     $*$Corresponding author: zuochao@njust.edu.cn; chenqian@njust.edu.cn
\end{affiliations}

\begin{abstract}
We report a computational 3D microscopy technique, termed Fourier ptychographic diffraction tomography (FPDT), that iteratively stitches together numerous variably illuminated, low-resolution images acquired with a low-numerical aperture (NA) objective in 3D Fourier space to create a wide field-of-view (FOV), high-resolution, depth-resolved complex refractive index (RI) image across large volumes. Unlike conventional optical diffraction tomography (ODT) approaches that rely on controlled bright-field illumination, holographic phase measurement, and high-NA objective detection, FPDT employs tomographic RI reconstruction from low-NA intensity-only measurements. In addition, FPDT incorporates high-angle dark-field illuminations beyond the NA of the objective, significantly expanding the accessible object frequency. With FPDT, we present the highest-throughput ODT results with 390$nm$ lateral resolution and 899$nm$ axial resolution across a 10$\times$  FOV of 1.77$mm^2$  and a depth of focus of $\sim$20$\mu m$. Billion-voxel 3D tomographic imaging results of biological samples establish FPDT as a powerful non-invasive and label-free tool for high-throughput 3D microscopy applications.
\end{abstract}

\section*{Introduction}
In optical microscopy, there has been a continued need to increase the resolution and contrast for visualizing 3D subcellular features of transparent biological samples over a large field-of-view (FOV) and an extended period of time. Confocal microscopy\cite{1,2}, as the paradigmatic tool of 3D microscopy, can collect serial optical sections from thick fluorescence specimens with both high resolution and high specificity. However, focused, high-intensity laser irradiation is detrimental to living cells and tissues\cite{3}. Two-photon\cite{4} and light-sheet microscopy\cite{5} are superior alternatives with advantages of deeper penetration, efficient light detection, and reduced photobleaching. More recently, super-resolution fluorescence microscopy has opened new views towards nanoscale subcellular structures\cite{6,7,8}. However, all these techniques require fluorescent proteins as biomarkers, and thus are ill-suited for samples that are non-fluorescent or cannot be fluorescently tagged. In addition, the photobleaching and phototoxicity of fluorescent agents prevent live cell imaging over extended periods of time\cite{9}. Overcoming these limitations becomes extremely challenging, as acquiring data over multi-dimension prolongs the exposure of the specimen to the excitation light, reducing its viability. Furthermore, the throughput of these imaging modalities is limited by the space–bandwidth product (SBP) of the optical system\cite{10}. In order to achieve sub-cellular imaging resolution and depth sectioning, a high-magnification objective has to be used, resulting in a proportionally smaller FOV. For example, a standard 60$\times$ objective with a 0.9 numerical aperture (NA) has a lateral resolution of $\sim$0.38$\mu m$ but a very limited FOV of $\sim$200$\mu m$ in diameter.

The refractive index (RI) distribution serves as an important endogenous contrast agent, which indeed enables the visualization of intracellular structures of biological samples without the need for specific staining or fluorescent tagging. In 1969, Wolf\cite{11} proposed optical diffraction tomography (ODT) as a solution to infer the 3D RI distribution by combining the X-ray tomography principle with optical holography. Different from conventional holography\cite{12,13} and other quantitative phase imaging (QPI) techniques\cite{14,15,16,17,18,19,20}, which only capture 2D integral phase shifts introduced by the sample, ODT is a true 3D imaging technique in the sense that the volumetric information inside the sample can be accessible\cite{21}. In a typical ODT system, the sample is illuminated from various directions\cite{22,23,24,25} or axially scanned at different depths\cite{26,27,28,29}, and the resulting complex diffraction patterns in the far field are measured based on QPI techniques. These measurements may then be synthesized in Fourier space with the Fourier slice theorem\cite{23,30} or Fourier diffraction theorem\cite{11,22} into a 3D tomographic reconstruction.

Due to synthetic aperture principles\cite{31}, ODT naturally has the additional benefit of increasing the effective imaging aperture to larger than (maximum of 2 times for the transmissive configuration) the physical aperture set by the microscope\cite{23,32}. Nevertheless, as a technique that generally requires both angular/depth scanning and multiple phase measurements, ODT has been primarily implemented in well-controlled, customized optical setups, prohibiting their widespread use in biological and medical science. Furthermore, to alleviate the ``missing cone" problem due to the limited angle of acceptance of the imaging system, ODT typically employs a high-magnification objective at the cost of significant FOV reduction\cite{33,34}, producing additional challenges for applications to several important problems such as rare cell imaging or optical phenotyping of model organisms, where large-scale high-throughput microscopy is highly desired.

Here, we present a computational ODT approach to achieve wide FOV, high-resolution, depth-resolved complex RI imaging across large volumes. Our technique, termed Fourier ptychographic diffraction tomography (FPDT), uses a low-NA objective to acquire a sequence of intensity images corresponding to different illumination angles scanned sequentially with a programmable light-emitting-diode (LED) array. Then, FPDT gradually combines these intensity images into a 3D spectrum of the object using an ODT-based ptychographic reconstruction algorithm. After the convergence of the algorithm, an inverse Fourier transform is performed to obtain the sample’s 3D RI distribution.

As its name implies, FPDT is inspired by Fourier ptychographic microscopy\cite{19,20} and Fourier ptychographic tomography\cite{35,36}, which overcome the physical SBP limit using angled illumination without mechanical scanning. In addition, since no explicit phase measurement is required, the corresponding system configuration eliminates the design challenges associated with interferometric detection schemes, and can be built upon an off-the-shelf microscope with the illumination unit replaced by a programmable LED array. However, our approach is not a simple extension of the state-of-the-arts as it first addresses the following three key issues regarding non-interferometic wide-field high-resolution ODT: (1) We establish a forward imaging model of FPDT that describes the measured intensity in terms of the scattered and unscattered fields for both bright-field and dark-field imaging in a precise and elegant way. The model can also accommodate both the first Born and Rytov approximations, and the latter has been proven to be more accurate for measuring thick biological samples. (2) In contrast to conventional ODT where only bright-field illumination is used, the synthetic aperture process of FTDT further incorporates high-angle dark-field illuminations that can be much greater than that allowed by the NA of the objective, which significantly expands the accessible object frequency. (3) We develop a FPDT platform can provide 1.3-NA resolution (390$nm$ lateral resolution and 899$nm$ axial resolution) across a 10$\times$ FOV of $\sim$1.77$mm^2$. As a result, our ODT platform combines high spatial resolution in 3D with a significantly large imaging volume, offering a 3D SBP that is unmatched by existing ODT approaches. Probing a large volume with a decent 3D spatial resolution, the FPDT could provide a powerful tool for high-throughput imaging applications in, e.g., cell and developmental biology.

\section*{Fourier diffraction theorem in finite-aperture optical systems}

\begin{figure}
  \centering
  \includegraphics[width={1.0\linewidth}]{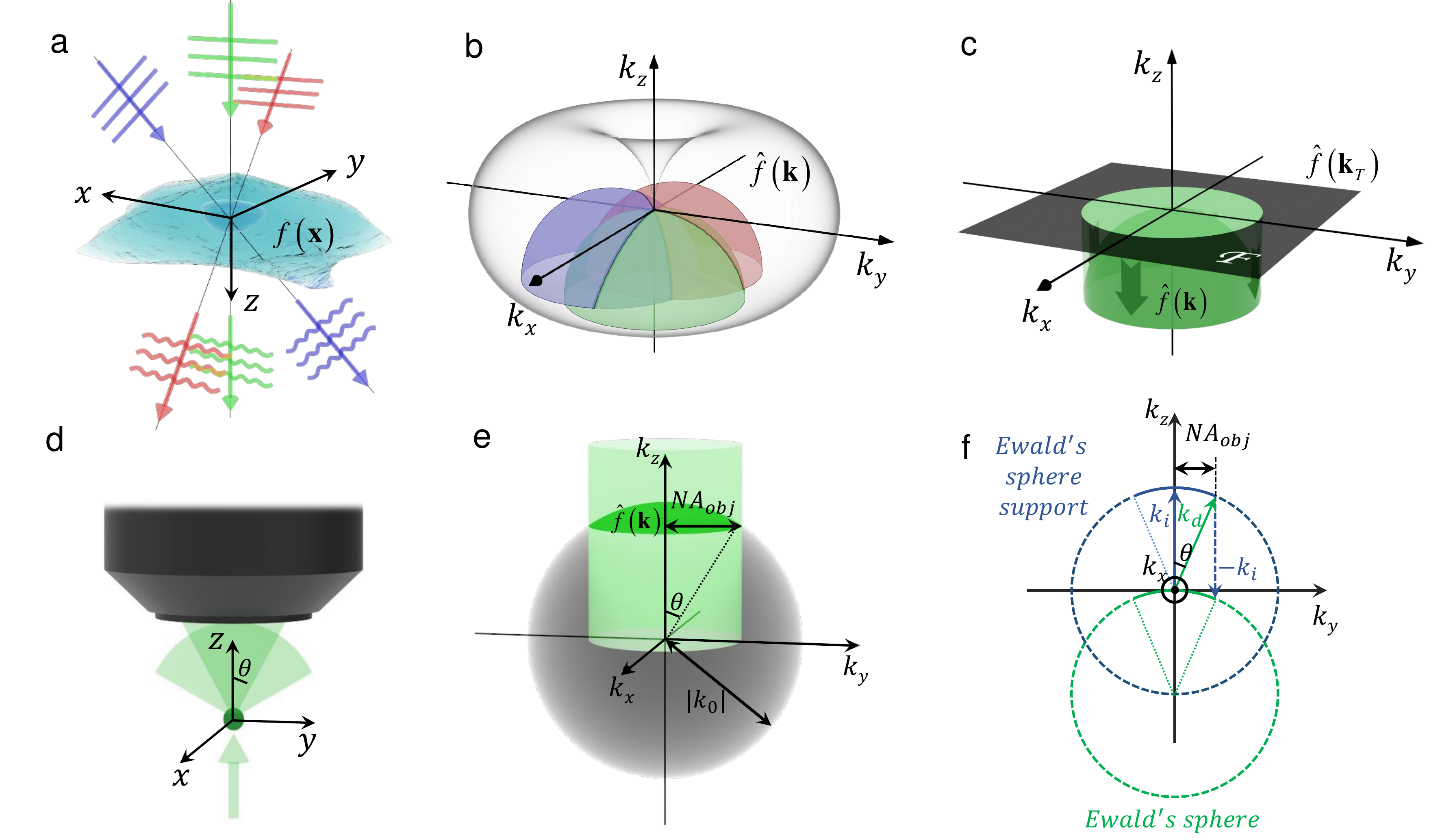}
  \caption{Fourier diffraction theorem in finite-aperture optical systems.
  a, A thick 3D sample is illuminated by plane waves from different directions.
  b, Particular semi-spherical surface in 3D Fourier space corresponding to each illumination direction.
  c, 2D Fourier spectrum projecting onto a semi-spherical surface.
  d, The forward scatted wave is limited by the aperture of the objective. e, f, 3D and 2D illustrations of the finite-aperture effect on the Fourier diffraction theorem.}
  \label{fig1}
\end{figure}

As illustrated in Fig.\,\ref{fig1}a, the physical quantity describing a thick 3D sample is termed as scattering potential $f{\left( {\bf{x}} \right)}$, which is a function of the 3D distribution of complex RI ${n\left( {\bf{x}} \right)}$ of the object:
\begin{equation}
f\left({\bf{x}}\right){\rm{  =  }}k_0^2\left[ {n{{\left( {\bf{x}} \right)}^2} - n_m^2} \right]
\label{eq1}
\end{equation}
where ${{\rm{ }}k_0 = {{2\pi } \mathord{\left/
 {\vphantom {{2\pi } \lambda }} \right.
 \kern-\nulldelimiterspace} \lambda }}$ is the wave-number in free space, $\lambda$ is the illumination wavelength, ${{n_m}}$ is the RI of the surrounding medium, and  ${{\bf{x}} \equiv \left( {x,y,z} \right) \equiv \left( {{{\bf{x}}_T},z} \right)}$ is a short-hand notation for the 3D spatial coordinate (transverse coordinate ${{{\bf{x}}_T}}$ and $z$). When a 3D sample is illuminated by a plane wave ${{U_{in}}\left( {\bf{x}} \right)}$, the resultant total field  ${U\left( {\bf{x}} \right)}$ is the superposition of the incident field, ${{U_{in}}\left( {\bf{x}} \right)}$, and the scattered field, ${{U_s}\left( {\bf{x}} \right)}$, that is, ${U\left( {\bf{x}} \right) = {U_{in}}\left( {\bf{x}} \right) + {U_s}\left( {\bf{x}} \right)}$.  To solve the inverse scattering problem, we introduce a new quantity ${{U_{s1}}({\bf{x}})}$, which represents the first-order scattered field. Under the (first-order)  Born\cite{11} or Rytov approximations\cite{22}, ${{U_{s1}}({\bf{x}})}$  is connected with ${{U_{in}}\left( {\bf{x}} \right)}$  and ${{U_s}\left( {\bf{x}} \right)}$  through the following equations (see \textbf{Supplementary Section A}):
\begin{equation}
 \begin{aligned}
{U_{s1}}({{\bf{x}}}) \approx \left\{ \begin{array}{l}
{U_s}({{\bf{x}}}) = U\left( {{{\bf{x}}}} \right) - {U_{in}}\left( {{{\bf{x}}}} \right){\kern 1pt} {\kern 1pt} {\kern 1pt} {\kern 1pt} {\kern 1pt} {\kern 1pt} {\kern 1pt} {\kern 1pt} {\kern 1pt} {\kern 1pt} {\kern 1pt} {\kern 1pt} {\kern 1pt} {\kern 1pt} {\kern 1pt} {\kern 1pt} {\kern 1pt} {\kern 1pt} {\kern 1pt} {\kern 1pt} {\kern 1pt} {\kern 1pt} {\kern 1pt} {\kern 1pt} {\kern 1pt} {\kern 1pt} {\kern 1pt} {\kern 1pt} {\kern 1pt} {\kern 1pt} {\kern 1pt} {\kern 1pt} {\kern 1pt} {\kern 1pt} {\kern 1pt} {\kern 1pt} {\kern 1pt} {\kern 1pt} {\kern 1pt}{\kern 1pt} {\kern 1pt} {\kern 1pt} {\kern 1pt} {\kern 1pt} {\kern 1pt} {\kern 1pt} {\kern 1pt} {\kern 1pt} {\kern 1pt} {\kern 1pt} {\kern 1pt} {\kern 1pt} {\kern 1pt} {\kern 1pt} {\kern 1pt} {\kern 1pt} {\kern 1pt} {\kern 1pt} {\kern 1pt} {\kern 1pt} {\kern 1pt} {\kern 1pt} {\kern 1pt} {\kern 1pt} {\kern 1pt} {\kern 1pt} {\kern 1pt} {\kern 1pt} {\kern 1pt} {\kern 1pt} {\kern 1pt} {\kern 1pt} {\kern 1pt} {\kern 1pt} {\kern 1pt} {\kern 1pt} {\kern 1pt} {\kern 1pt} {\kern 1pt} {\kern 1pt} {\kern 1pt} {\kern 1pt} {\kern 1pt} {\kern 1pt} {\kern 1pt} {\kern 1pt} {\kern 1pt} {\kern 1pt} {\kern 1pt} {\kern 1pt} {\kern 1pt} {\kern 1pt} {\kern 1pt} {\kern 1pt} {\kern 1pt} {\kern 1pt} {\kern 1pt} {\kern 1pt} {\kern 1pt} {\kern 1pt} {\kern 1pt} {\kern 1pt} {\kern 1pt} {\kern 1pt} {\kern 1pt} {\kern 1pt} {\kern 1pt} {\kern 1pt} {\kern 1pt} {\kern 1pt} {\kern 1pt} {\kern 1pt} {\kern 1pt} {\kern 1pt} {\kern 1pt} {\kern 1pt} {\kern 1pt} {\kern 1pt} {\kern 1pt} {\kern 1pt} {\kern 1pt} {\kern 1pt} {\kern 1pt} Born{\kern 1pt} {\kern 1pt} {\kern 1pt} {\kern 1pt} approximation\\
{U_{in}}({{\bf{x}}})\ln \left( {\frac{{U\left( {{{\bf{x}}}} \right)}}{{{U_{in}}\left( {{{\bf{x}}}} \right)}}} \right){\kern 1pt} {\kern 1pt}  = {\kern 1pt} {\kern 1pt} {U_{in}}({{\bf{x}}})\ln \left( {\frac{{{U_s}\left( {{{\bf{x}}}} \right) + {U_{in}}\left( {{{\bf{x}}}} \right)}}{{{U_{in}}\left( {{{\bf{x}}}} \right)}}} \right) {\kern 1pt}{\kern 1pt} {\kern 1pt} {\kern 1pt} {\kern 1pt}  {\kern 1pt}  {\kern 1pt}  {\kern 1pt}  {\kern 1pt}  {\kern 1pt}  {\kern 1pt}  {\kern 1pt}  {\kern 1pt}  {\kern 1pt}  {\kern 1pt}  {\kern 1pt} {\kern 1pt}Rytov{\kern 1pt} {\kern 1pt} {\kern 1pt} approximation
\end{array} \right.
\label{eq2}
 \end{aligned}
\end{equation}
Based on Green's function method\cite{37}, the linearized relation between the first-order scattered field and the scattering potential of the object can be established, and the corresponding form in the frequency domain is well-known as the Fourier diffraction theorem\cite{11}:
\begin{equation}
\hat f\left( {{\bf{k}} - {{\bf{k}}_i}} \right) =  - \frac{{\exp \left( { - j{k_z}{z_D}} \right)j{k_z}}}{\pi }{\hat U_{s1}}({{\bf{k}}_T};z = {z_D})P\left( {{{\bf{k}}_T}} \right)\delta \left( {{k_z} - \sqrt {k_m^2 - {{\left| {{{\bf{k}}_T}} \right|}^2}} } \right)
\label{eq3}
\end{equation}
where $j$ is the imaginary unit, $k_m$ is the wave-number in the medium, ${{{\bf{k}}_i}}$ is the 3D wave vector of the incident plane wave, the exponential term in Fig.\,\ref{eq3} accounts for the coordinate shift in the $z$ direction and will automatically vanish if the measurement is performed at the nominal ‘in-focus’ plane (${z_D = 0}$), and ${\hat f\left( {\bf{k}} \right)}$ and ${{\hat U_{s1}}({{\bf{k}}_T};z = {z_D})}$ are the 3D and 2D Fourier transforms of ${f\left( {\bf{x}} \right)}$ and ${{U_{s1}}\left( {{{\bf{x}}_T};z = {z_D}} \right)}$, respectively (we use the ``hat" to denote the signal spectrum in the 2D/3D Fourier domain). Because the 3D frequency vector, ${{\bf{k}} = \left( {{{\bf{k}}_T},{k_z}} \right)}$, lies on the 2D surface of the so-called Ewald sphere under the constraint ${{k_z} = \sqrt {k_m^2 - {{\left| {{{\bf{k}}_T}} \right|}^2}}}$, the information defined by ${{\hat U_{s1}}({{\bf{k}}_T})}$, is directly related to a particular semi-spherical surface with a radius of ${k_m^{}}$ in 3D Fourier space that is displaced by ${-{{\bf{k}}_i}}$ (see Fig. 1b). Thus, the planar 2D Fourier spectrum is projected onto a semi-spherical surface, as depicted in Fig. 1c. However, for a practical microscopic system, only forward propagating waves falling within the system aperture can contribute to the image formation, as illustrated in Fig. 1d. In 2D imaging, the effect of the lens aperture is usually described by the 2D complex pupil function [i.e., coherent transfer function (CTF)] ${P\left({{{\bf{k}}_T}} \right)}$, which ideally is a circ-function with a radius of ${{k_0}{NA}{_{obj}}}$, determined by the NA of the objective. For 3D imaging, the complex pupil function should be projected onto the spherical surface, i.e., ${P\left( {\bf{k}} \right) = P\left( {{{\bf{k}}_T}} \right)\delta \left( {{k_z} - \sqrt {k_m^2 - {{\left| {{{\bf{k}}_T}} \right|}^2}} } \right)}$, resulting a subsection of the Ewald sphere called the generalized aperture (i.e., 3D CTF)\cite{38} (see Fig.\,\ref{fig1}e). Limited by the generalized aperture, the forward scattered field by sub-wavelength delta-like features (subtending an angle of ±90\degree) cannot be completely captured by the microscope. As shown in Fig.\,\ref{fig1}e,f, the aperture angle $\theta$ denotes the largest cone of wave-vectors that can pass from the sample into the imaging lens, so the generalized aperture can cover the half-sphere only when $\theta$ = 90\degree, .

The Fourier diffraction theorem (Fig.\,\ref{eq3}) suggests that for each illumination angle, only partial spherical cap bounded by the generalized aperture can be probed. Illuminating the object at different angles will shift different regions of the object’s frequency spectrum into a fixed microscope objective lens, enlarging the accessible object frequency domain. In conventional ODT systems, the complex amplitude (both amplitude and phase) of the total field, ${U\left({{{\bf{x}}_T}} \right)}$, is measured by interferometric or holographic approaches, and the complex function of incident plane wave illumination, ${{U_{in}}\left({\bf{x}} \right)}$, can be known beforehand or determined through proper calibration. Thus, ${{U_{{\rm{s1}}}}({{\bf{x}}_T})}$ can be calculated by Fig.\,\ref{eq2} with either first Born approximation or Rytov approximation, and then mapped on the particular Eward sphere according to Fig.\,\ref{eq3}. With a series of angle-dependent interferometric complex amplitude measurements, certain regions of the 3D Fourier spectrum of the object can be completed, which will allow us to reconstruct the scattering potential of the 3D sample. It should be mentioned that the Rytov approximation is valid as long as the phase gradient in the sample is small and is independent of the sample size and the total phase shift, so it has been considered to be less restrictive and has been shown to lead to a reconstruction that is superior to that of the Born approximation for thick biological samples\cite{22}.

\section*{FPDT forward imaging model and reconstruction algorithm}
Since all computational imaging approaches rely on inversion algorithms to retrieve the parameters of interest of the sample, they require an accurate modelling of the link between the measured data and the sample parameters. For a thin sample characterized by its 2D amplitude and phase, each angled illumination shifts the object spectrum around in 2D Fourier space, with the objective aperture selecting out different sections\cite{19}. The recorded images are the intensities of resultant complex fields corresponding to different segments of the object spectrum. For a thick sample characterized by its 3D scattering potential, the object spectrum is shifted by the incident wave-vector in 3D Fourier space in a similar manner, and only the frequencies covered by the 3D generalized aperture can pass through the imaging system and contribute to the image formation. However, ${U_{s1}}({\bf{x}})$ is not the field detected in the image plane. Instead, it is related to our measurements by Fig.\,\ref{eq2} with either the Born or Rytov approximation. Therefore, a forward imaging model linking the measured intensities with the sample's scattering potential should be established.
\begin{figure}
  \centering
  \includegraphics[width={1.0\linewidth}]{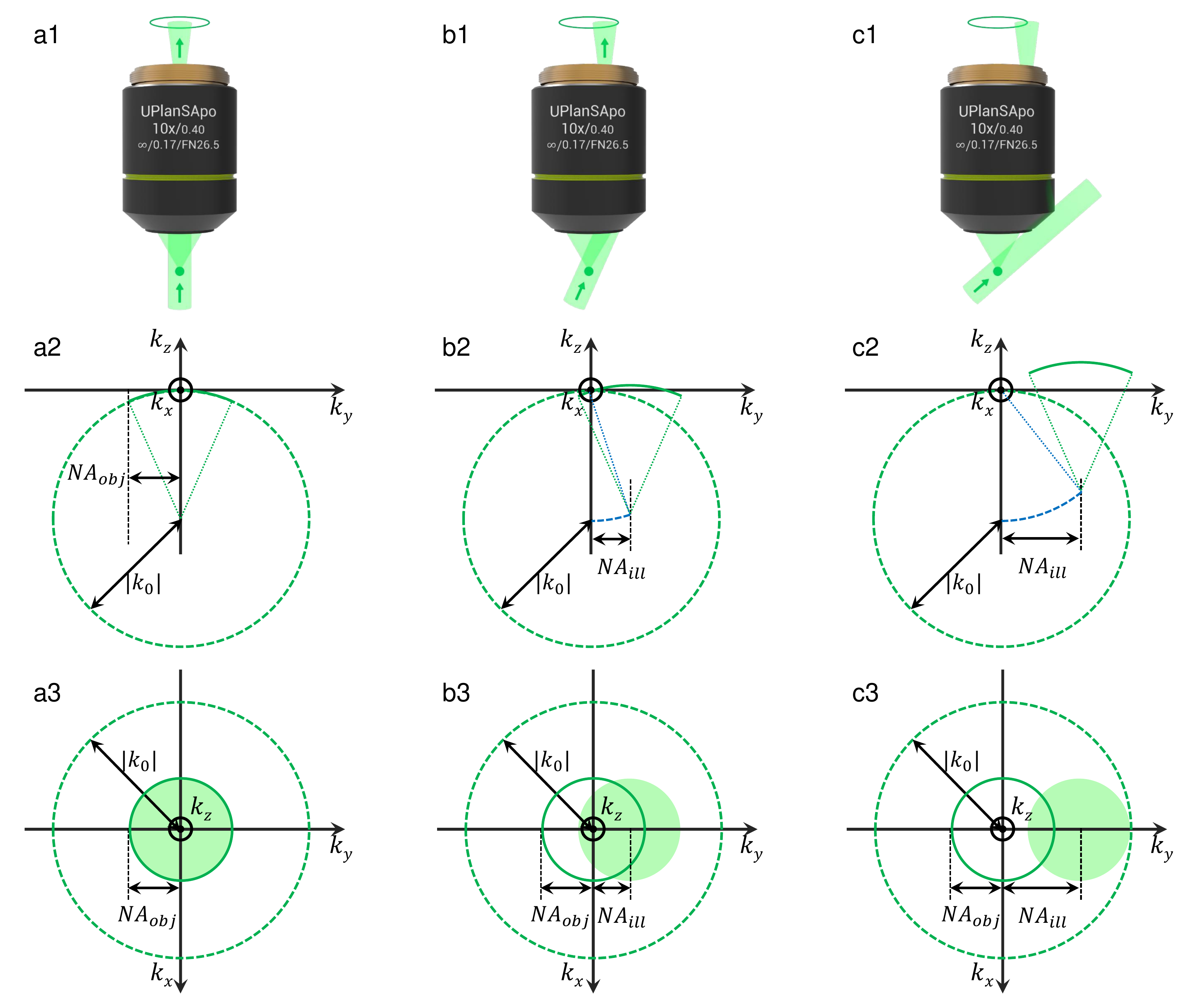}
  \caption{Forward imaging model of FPDT for different incident angles.
  a, Bright-field imaging with on-axis illumination.
  b, Bright-field imaging with tilted illumination.
  c, Dark-field imaging with large-angle illumination.
  a1-c1, Schematics illustrating the image formation of FPDT under different illumination conditions.
  a2-c3, 2D spectral support diagrams in the ky-kz and kx-ky planes.}
  \label{fig2}
\end{figure}

As illustrated in Fig.\,\ref{fig2}a,b, for bright-field imaging where the incident illumination falls within the objective pupil, the field measured is the total field ${U\left( {\bf{x}} \right)}$ [the sum of ${{U_{in}}\left( {\bf{x}} \right)}$ and ${{U_s}\left( {\bf{x}} \right)}$], which is similar to the case of conventional ODT but only the intensity is recorded. For dark-field imaging illustrated in Fig.\,\ref{fig2}c, the 3D pupil function does not intersect the zero order of the object spectrum at ${{\bf{k}} = {\bf{0}}}$, and the incident illumination (zero diffraction order), which falls out of the objective, is absent in the image. Hence, only the scattered component contributes to the image formation.
\begin{equation}
I({{\bf{x}}_T}) = \left\{ \begin{array}{l}
{\left| {U\left( {{{\bf{x}}_T}} \right)} \right|^2}={\left| {{U_s}\left( {{{\bf{x}}_T}} \right) + {U_{in}}\left( {{{\bf{x}}_T}} \right)} \right|^2}{\kern 1pt} {\kern 1pt} {\kern 1pt} {\kern 1pt} {\kern 1pt} {\kern 1pt} {\kern 1pt} {\kern 1pt} {\kern 1pt} {\kern 1pt} {\kern 1pt} {\kern 1pt} {\kern 1pt} {\kern 1pt} {\kern 1pt} {\kern 1pt} {\kern 1pt} {\kern 1pt} Bright{\kern 1pt} field{\kern 1pt} \\
{\left| {{U_s}\left( {{{\bf{x}}_T}} \right)} \right|^2}{\kern 1pt} {\kern 1pt} {\kern 1pt} {\kern 1pt} {\kern 1pt} {\kern 1pt} {\kern 1pt} {\kern 1pt} {\kern 1pt} {\kern 1pt} {\kern 1pt} {\kern 1pt} {\kern 1pt} {\kern 1pt} {\kern 1pt} {\kern 1pt} {\kern 1pt} {\kern 1pt} {\kern 1pt} {\kern 1pt} {\kern 1pt} {\kern 1pt} {\kern 1pt} {\kern 1pt} {\kern 1pt} {\kern 1pt} {\kern 1pt} {\kern 1pt} {\kern 1pt} {\kern 1pt} {\kern 1pt} {\kern 1pt} {\kern 1pt} {\kern 1pt} {\kern 1pt} {\kern 1pt} {\kern 1pt} {\kern 1pt} {\kern 1pt} {\kern 1pt} {\kern 1pt} {\kern 1pt} {\kern 1pt} {\kern 1pt} {\kern 1pt} {\kern 1pt} {\kern 1pt} {\kern 1pt} {\kern 1pt} {\kern 1pt} {\kern 1pt} {\kern 1pt} {\kern 1pt} {\kern 1pt} {\kern 1pt} {\kern 1pt} {\kern 1pt} {\kern 1pt} {\kern 1pt} {\kern 1pt} {\kern 1pt} {\kern 1pt} {\kern 1pt} {\kern 1pt} {\kern 1pt} {\kern 1pt} {\kern 1pt} {\kern 1pt} {\kern 1pt} {\kern 1pt} {\kern 1pt} {\kern 1pt} {\kern 1pt} {\kern 1pt} {\kern 1pt} {\kern 1pt} {\kern 1pt} {\kern 1pt} {\kern 1pt} {\kern 1pt} {\kern 1pt} {\kern 1pt} {\kern 1pt} {\kern 1pt} {\kern 1pt} {\kern 1pt} {\kern 1pt} {\kern 1pt} {\kern 1pt} {\kern 1pt} {\kern 1pt} {\kern 1pt} {\kern 1pt} {\kern 1pt} {\kern 1pt} {\kern 1pt} {\kern 1pt} {\kern 1pt} {\kern 1pt} {\kern 1pt} {\kern 1pt} {\kern 1pt} {\kern 1pt} {\kern 1pt} {\kern 1pt} {\kern 1pt} {\kern 1pt} {\kern 1pt} {\kern 1pt} {\kern 1pt} {\kern 1pt} {\kern 1pt} {\kern 1pt} {\kern 1pt} {\kern 1pt} {\kern 1pt} {\kern 1pt} {\kern 1pt} {\kern 1pt} {\kern 1pt} {\kern 1pt} {\kern 1pt} {\kern 1pt} {\kern 1pt}{\kern 1pt}{\kern 1pt}{\kern 1pt}{\kern 1pt}{\kern 1pt}{\kern 1pt} {\kern 1pt} {\kern 1pt} {\kern 1pt} {\kern 1pt} {\kern 1pt} Dark{\kern 1pt} {\kern 1pt} field{\kern 1pt} {\kern 1pt} {\kern 1pt} {\kern 1pt} {\kern 1pt} {\kern 1pt} {\kern 1pt} {\kern 1pt}
\end{array} \right.
\label{eq4}
\end{equation}
In the following text, we introduce three new quantities, ${U_{sn}^{}({{\bf{x}}_T})}$, ${U_n^{}({{\bf{x}}_T})}$  and ${U_{s1n}^{}({{\bf{x}}_T})}$, which are normalized versions of the first-order scattered field, total field, and scattered field with respect to the incident field respectively, i,e,, ${{{U_{sn}^{}({{\bf{x}}_T}) = U_s^{}({{\bf{x}}_T})} \mathord{\left/
 {\vphantom {{U_{sn}^{}({{\bf{x}}_T}) = U_s^{}({{\bf{x}}_T})} {U_{in}^{}({{\bf{x}}_T})}}} \right.
 \kern-\nulldelimiterspace} {U_{in}^{}({{\bf{x}}_T})}}}$, ${{{U_n^{}({{\bf{x}}_T}) = U({{\bf{x}}_T})} \mathord{\left/
 {\vphantom {{U_n^{}({{\bf{x}}_T}) = U({{\bf{x}}_T})} {U_{in}^{}({{\bf{x}}_T})}}} \right.
 \kern-\nulldelimiterspace} {U_{in}^{}({{\bf{x}}_T})}}}$, and ${{{U_{s1n}^{}({{\bf{x}}_T}) = U_{s1}^{}({{\bf{x}}_T})} \mathord{\left/
 {\vphantom {{U_{s1n}^{}({{\bf{x}}_T}) = U_{s1}^{}({{\bf{x}}_T})} {U_{in}^{}({{\bf{x}}_T})}}} \right.
 \kern-\nulldelimiterspace} {U_{in}^{}({{\bf{x}}_T})}}}$. Without loss of generality, the incident illumination is considered to be an angled plane wave with a unit amplitude, so the normalization process will not modify the intensities of the detected fields. With these new notations, Fig.\,\ref{eq2} can be simplified as follows:
\begin{equation}
{U_{s1n}}({{\bf{x}}_T}) = \left\{ \begin{array}{l}
{U_{sn}}({{\bf{x}}_T}) = {U_n}\left( {{{\bf{x}}_T}} \right) - 1{\kern 1pt} {\kern 1pt} {\kern 1pt} {\kern 1pt} {\kern 1pt} {\kern 1pt} {\kern 1pt} {\kern 1pt} {\kern 1pt} {\kern 1pt} {\kern 1pt} {\kern 1pt} {\kern 1pt} {\kern 1pt} {\kern 1pt} {\kern 1pt} {\kern 1pt} {\kern 1pt} {\kern 1pt} {\kern 1pt} {\kern 1pt} {\kern 1pt} {\kern 1pt} {\kern 1pt} {\kern 1pt} {\kern 1pt} {\kern 1pt} {\kern 1pt} {\kern 1pt} {\kern 1pt} {\kern 1pt} {\kern 1pt} {\kern 1pt} {\kern 1pt} {\kern 1pt} {\kern 1pt} {\kern 1pt} {\kern 1pt} {\kern 1pt} {\kern 1pt} {\kern 1pt} {\kern 1pt} {\kern 1pt} {\kern 1pt} {\kern 1pt} {\kern 1pt} {\kern 1pt} {\kern 1pt} {\kern 1pt} {\kern 1pt} {\kern 1pt} {\kern 1pt} {\kern 1pt} {\kern 1pt} {\kern 1pt} {\kern 1pt} {\kern 1pt} {\kern 1pt} {\kern 1pt} {\kern 1pt} {\kern 1pt} {\kern 1pt} {\kern 1pt} {\kern 1pt} {\kern 1pt} {\kern 1pt} {\kern 1pt} {\kern 1pt} {\kern 1pt} {\kern 1pt} {\kern 1pt} {\kern 1pt} {\kern 1pt} {\kern 1pt} {\kern 1pt} {\kern 1pt} {\kern 1pt} Born{\kern 1pt} {\kern 1pt} {\kern 1pt} {\kern 1pt} approximation\\
\ln \left( {{U_n}({{\bf{x}}_T})} \right){\kern 1pt} {\kern 1pt} {\kern 1pt}  = {\kern 1pt} {\kern 1pt} {\kern 1pt} \ln \left( {{U_{sn}}({{\bf{x}}_T}) + 1} \right){\kern 1pt} {\kern 1pt} {\kern 1pt} {\kern 1pt} {\kern 1pt} {\kern 1pt} {\kern 1pt} {\kern 1pt} {\kern 1pt} {\kern 1pt}{\kern 1pt}  {\kern 1pt} {\kern 1pt} {\kern 1pt} {\kern 1pt} {\kern 1pt} {\kern 1pt} {\kern 1pt} {\kern 1pt} {\kern 1pt} {\kern 1pt} {\kern 1pt} {\kern 1pt} {\kern 1pt} Rytov{\kern 1pt} {\kern 1pt} {\kern 1pt} approximation
\end{array} \right.
\label{eq5}
\end{equation}
The inverse relation, which connects the measured field to the first-order scattered field (scattered potential) is
\begin{equation}
\left\{ \begin{array}{l}
{U_n}({{\bf{x}}_T}) = {U_{s1n}}({{\bf{x}}_T}) + 1{\kern 1pt} {\kern 1pt} {\kern 1pt} {\kern 1pt} {\kern 1pt} {\kern 1pt} {\kern 1pt} {\kern 1pt} {\kern 1pt} {\kern 1pt} {\kern 1pt} {\kern 1pt} {\kern 1pt} {\kern 1pt} {\kern 1pt} {\kern 1pt} {\kern 1pt} {\kern 1pt} {\kern 1pt} {\kern 1pt} {\kern 1pt} {\kern 1pt} {\kern 1pt} {\kern 1pt} {\kern 1pt} {\kern 1pt} {\kern 1pt} {\kern 1pt} {\kern 1pt} {\kern 1pt} {\kern 1pt} {\kern 1pt} {\kern 1pt} {\kern 1pt} {\kern 1pt} {\kern 1pt} {\kern 1pt} {\kern 1pt} {\kern 1pt} {\kern 1pt} {\kern 1pt} {\kern 1pt} {\kern 1pt} {\kern 1pt} {\kern 1pt} {\kern 1pt} {\kern 1pt} Bright{\kern 1pt} field\\
{U_{sn}}({{\bf{x}}_T}) = {U_{s1n}}\left( {{{\bf{x}}_T}} \right){\kern 1pt} {\kern 1pt} {\kern 1pt} {\kern 1pt} {\kern 1pt} {\kern 1pt} {\kern 1pt} {\kern 1pt} {\kern 1pt} {\kern 1pt} {\kern 1pt} {\kern 1pt} {\kern 1pt} {\kern 1pt} {\kern 1pt} {\kern 1pt} {\kern 1pt} {\kern 1pt} {\kern 1pt} {\kern 1pt} {\kern 1pt} {\kern 1pt} {\kern 1pt} {\kern 1pt} {\kern 1pt} {\kern 1pt} {\kern 1pt} {\kern 1pt} {\kern 1pt} {\kern 1pt} {\kern 1pt} {\kern 1pt} {\kern 1pt} {\kern 1pt} {\kern 1pt} {\kern 1pt} {\kern 1pt} {\kern 1pt} {\kern 1pt} {\kern 1pt} {\kern 1pt} {\kern 1pt} {\kern 1pt} {\kern 1pt} {\kern 1pt} {\kern 1pt} {\kern 1pt} {\kern 1pt} {\kern 1pt} {\kern 1pt} {\kern 1pt} {\kern 1pt} {\kern 1pt} {\kern 1pt} {\kern 1pt} {\kern 1pt} {\kern 1pt} {\kern 1pt} Dark{\kern 1pt} {\kern 1pt} field{\kern 1pt}
\end{array} \right.
\label{eq6}
\end{equation}
for the Born approximation, and
\begin{equation}
\left\{ \begin{array}{l}
{U_n}({{\bf{x}}_T}) = \exp \left( {{U_{s1n}}({{\bf{x}}_T})} \right){\kern 1pt} {\kern 1pt} {\kern 1pt} {\kern 1pt} {\kern 1pt} {\kern 1pt} {\kern 1pt} {\kern 1pt} {\kern 1pt} {\kern 1pt} {\kern 1pt} {\kern 1pt} {\kern 1pt} {\kern 1pt} {\kern 1pt} {\kern 1pt} {\kern 1pt} {\kern 1pt} {\kern 1pt} {\kern 1pt} {\kern 1pt} {\kern 1pt} {\kern 1pt} {\kern 1pt} {\kern 1pt} {\kern 1pt} {\kern 1pt} {\kern 1pt} {\kern 1pt} {\kern 1pt} {\kern 1pt} {\kern 1pt} {\kern 1pt} {\kern 1pt} {\kern 1pt} {\kern 1pt} {\kern 1pt} {\kern 1pt} {\kern 1pt} {\kern 1pt} {\kern 1pt} {\kern 1pt} {\kern 1pt} {\kern 1pt} {\kern 1pt} {\kern 1pt} {\kern 1pt} {\kern 1pt} {\kern 1pt} {\kern 1pt} {\kern 1pt} {\kern 1pt} {\kern 1pt} {\kern 1pt} {\kern 1pt} {\kern 1pt} {\kern 1pt} {\kern 1pt} {\kern 1pt} {\kern 1pt}   Bright{\kern 1pt} field\\
{U_{sn}}({{\bf{x}}_T}) = \exp \left( {{U_{s1n}}({{\bf{x}}_T})} \right) - 1{\kern 1pt} {\kern 1pt} {\kern 1pt} {\kern 1pt} {\kern 1pt} {\kern 1pt} {\kern 1pt} {\kern 1pt} {\kern 1pt} {\kern 1pt} {\kern 1pt} {\kern 1pt} {\kern 1pt} {\kern 1pt} {\kern 1pt} {\kern 1pt} {\kern 1pt} {\kern 1pt} {\kern 1pt} {\kern 1pt} {\kern 1pt} {\kern 1pt} {\kern 1pt} {\kern 1pt} {\kern 1pt} {\kern 1pt} {\kern 1pt} {\kern 1pt} {\kern 1pt} {\kern 1pt} {\kern 1pt} {\kern 1pt} {\kern 1pt} {\kern 1pt} {\kern 1pt}{\kern 1pt}  {\kern 1pt} Dark{\kern 1pt} {\kern 1pt} field
\end{array} \right.
\label{eq7}
\end{equation}
for the Rytov approximation. Based on these imaging models, we develop an iterative reconstruction algorithm, mirroring that from FPM\cite{19}, to ``fill in" the high-resolution k-space scattering potential computationally with data from a set of $N$ captured low resolution intensity images, ${I_c^i\left( {{\bf{x}}_T^{}} \right)}$, with their corresponding illumination wavevector ${{\bf{k}}_{in}^i}$, with ${i = 1,2,...,N}$. Here we use the superscript $i$ to denote that this image is formed by the ${i}$th LED illumination. Figure 3a displays the general process of the FPDT reconstruction algorithm, which alternates between the spatial and Fourier domain according to the following steps:
\begin{enumerate}
 \item  Make an initial guess of the high-resolution k-space scattering potential, ${\hat f\left({\bf{k}} \right)}$. It has been found that the initial value for ${\hat f\left( {\bf{k}} \right)}$ is not critical to the final reconstruction result, so we simply initialize ${\hat f\left( {\bf{k}} \right)}$ with all zeros in this work.
 \item From the first illumination angle, select values of ${\hat f\left( {\bf{k}} \right)}$ taken along its associated shell corresponding to ${{\bf{k}} - {{\bf{k}}_i}}$  and bounded by the 3D generalized aperture, ${P\left( {\bf{k}} \right)}$  (radius of ${k_0^{}}$, and maximum width of ${2k_0^{}{NA}{_{obj}}}$). The sub-region of the 3D spectrum is projected along the axial frequency coordinate to obtain a low-resolution 2D Fourier sub-spectrum ${\hat f_{}^i\left( {{\bf{k}}_T^{}} \right)}$, which is directly related to ${U_{s1}^i({{\bf{x}}_T})}$ after correcting for some constants according to ${\hat U_{s1}^i({{\bf{k}}_T}) = -\frac{{\pi j}}{{{k_z}}}\hat f_{}^i\left( {{\bf{k}}_T^{}} \right)}$.
 \item Circular shift ${\hat U_{s1}^i({{\bf{k}}_T})}$  in the frequency domain with a translation vector ${- {\bf{k}}_{inT}^i =  - (k_{inx{\kern 1pt} }^i,k_{iny{\kern 1pt} }^i)}$, and inverse Fourier transform the resultant spectrum ${{\hat U_{s1}}({{\bf{k}}_T} - {\bf{k}}_{inT}^i)}$  to the spatial domain. This is equivalent to the actual physical scenario of using our current high-resolution k-space scattering potential estimate to simulate the formation of the low-resolution normalized first-order scattered field, ${{{U_{s1n}^i({{\bf{x}}_T}) = U_s^i({{\bf{x}}_T})} \mathord{\left/
 {\vphantom {{U_{s1n}^i({{\bf{x}}_T}) = U_s^i({{\bf{x}}_T})} {U_{in}^i({{\bf{x}}_T})}}} \right.
 \kern-\nulldelimiterspace} {U_{in}^i({{\bf{x}}_T})}}}$, for the  ${i}$th illumination angle. As the 2D complex amplitude of the incident illumination only contains a phase ramp depending on the tilt angle, i.e., ${U_{in}^i\left( {{{\bf{x}}_T}} \right) = {e^{j{\bf{k}}_{inT}^i \cdot {{\bf{x}}_T}}}}$, the Fourier shift property suggests that multiplying or dividing by ${U_{in}^i\left( {{{\bf{x}}_T}} \right)}$  in the spatial domain is equivalent to a circular shift of the signal spectrum by ${\mp {\bf{k}}_{inT}^i =  \mp (k_{{\kern 1pt} inx}^i,k_{iny}^i)}$  in the frequency domain.
 \item Convert the estimated ${U_{s1n}^i({{\bf{x}}_T})}$  to the measured field and enforce the amplitude constraint. Note that the update formula depends both on the area which the illumination vector belongs to (bright-field or dark-field) and the approximation assumed for ODT (Born or Rytvo). For Born approximation, the update formula is derived based on the relation given by Eqs.\,\ref{eq5} and \ref{eq6}
\begin{equation}
\bar U_{s1n}^i({{\bf{x}}_T}) \approx \left\{ \begin{array}{l}
\sqrt {I_c^i\left( {{\bf{x}}_T^{}} \right)} \frac{{\exp \left( {U_{s1n}^i({{\bf{x}}_T}) + 1} \right){\kern 1pt} }}{{\left| {\exp \left( {U_{s1n}^i({{\bf{x}}_T}) + 1} \right)} \right|}}{\kern 1pt} {\kern 1pt}  - 1{\kern 1pt} {\kern 1pt} {\kern 1pt} {\kern 1pt} {\kern 1pt} {\kern 1pt} {\kern 1pt} {\kern 1pt} {\kern 1pt} {\kern 1pt} {\kern 1pt} {\kern 1pt} {\kern 1pt} {\kern 1pt} {\kern 1pt} {\kern 1pt} {\kern 1pt} {\kern 1pt} {\kern 1pt} {\kern 1pt} {\kern 1pt} {\kern 1pt} {\kern 1pt} {\kern 1pt} {\kern 1pt} {\kern 1pt} {\kern 1pt} {\kern 1pt} {\kern 1pt} {\kern 1pt} {\kern 1pt} {\kern 1pt} {\kern 1pt} {\kern 1pt} {\kern 1pt} {\kern 1pt} {\kern 1pt} {\kern 1pt} {\kern 1pt} {\kern 1pt} {\kern 1pt} {\kern 1pt} {\kern 1pt} {\kern 1pt} {\kern 1pt} {\kern 1pt} {\kern 1pt} {\kern 1pt} {\kern 1pt} {\kern 1pt} {\kern 1pt} {\kern 1pt} {\kern 1pt} {\kern 1pt} {\kern 1pt} {\kern 1pt} {\kern 1pt} {\kern 1pt} {\kern 1pt} {\kern 1pt} {\kern 1pt} {\kern 1pt} {\kern 1pt} {\kern 1pt} {\kern 1pt} {\kern 1pt} {\kern 1pt} {\kern 1pt} {\kern 1pt} {\kern 1pt} {\kern 1pt} {\kern 1pt} {\kern 1pt} {\kern 1pt} {\kern 1pt} {\kern 1pt} {\kern 1pt} {\kern 1pt} {\kern 1pt} {\kern 1pt} {\kern 1pt} {\kern 1pt} {\kern 1pt} {\kern 1pt} {\kern 1pt} {\kern 1pt}  {\kern 1pt}{\kern 1pt} {\kern 1pt} {\kern 1pt} {\kern 1pt} {\kern 1pt} {\kern 1pt} {\kern 1pt}  Bright{\kern 1pt} field\\
\sqrt {I_c^i\left( {{\bf{x}}_T^{}} \right)} \frac{{\exp \left( {U_{s1n}^i({{\bf{x}}_T})} \right){\kern 1pt} }}{{\left| {\exp \left( {U_{s1n}^i({{\bf{x}}_T})} \right)} \right|}}{\kern 1pt} {\kern 1pt} {\kern 1pt} {\kern 1pt} {\kern 1pt} {\kern 1pt} {\kern 1pt} {\kern 1pt} {\kern 1pt} {\kern 1pt} {\kern 1pt} {\kern 1pt} {\kern 1pt} {\kern 1pt} {\kern 1pt} {\kern 1pt} {\kern 1pt} {\kern 1pt} {\kern 1pt} {\kern 1pt} {\kern 1pt} {\kern 1pt} {\kern 1pt} {\kern 1pt} {\kern 1pt} {\kern 1pt} {\kern 1pt} {\kern 1pt} {\kern 1pt} {\kern 1pt} {\kern 1pt} {\kern 1pt} {\kern 1pt} {\kern 1pt} {\kern 1pt} {\kern 1pt} {\kern 1pt} {\kern 1pt} {\kern 1pt} {\kern 1pt} {\kern 1pt} {\kern 1pt} {\kern 1pt} {\kern 1pt} {\kern 1pt} {\kern 1pt} {\kern 1pt} {\kern 1pt} {\kern 1pt} {\kern 1pt} {\kern 1pt} {\kern 1pt} {\kern 1pt} {\kern 1pt} {\kern 1pt} {\kern 1pt} {\kern 1pt} {\kern 1pt} {\kern 1pt} {\kern 1pt} {\kern 1pt} {\kern 1pt} {\kern 1pt} {\kern 1pt} {\kern 1pt} {\kern 1pt} {\kern 1pt} {\kern 1pt} {\kern 1pt} {\kern 1pt}  {\kern 1pt} {\kern 1pt}{\kern 1pt} {\kern 1pt} {\kern 1pt} {\kern 1pt} {\kern 1pt} {\kern 1pt} {\kern 1pt} {\kern 1pt} {\kern 1pt} {\kern 1pt} {\kern 1pt} {\kern 1pt} {\kern 1pt} {\kern 1pt} {\kern 1pt} {\kern 1pt} {\kern 1pt} {\kern 1pt} {\kern 1pt} {\kern 1pt} {\kern 1pt} {\kern 1pt} {\kern 1pt} {\kern 1pt} {\kern 1pt} {\kern 1pt} {\kern 1pt} {\kern 1pt} {\kern 1pt} {\kern 1pt} {\kern 1pt} {\kern 1pt} {\kern 1pt} {\kern 1pt} {\kern 1pt} {\kern 1pt} {\kern 1pt} {\kern 1pt} {\kern 1pt} {\kern 1pt} {\kern 1pt} {\kern 1pt} {\kern 1pt} {\kern 1pt} {\kern 1pt} {\kern 1pt} {\kern 1pt} {\kern 1pt} {\kern 1pt} {\kern 1pt} {\kern 1pt} {\kern 1pt} {\kern 1pt} {\kern 1pt} {\kern 1pt} {\kern 1pt} Dark{\kern 1pt} field
\end{array} \right.
\label{eq8}
\end{equation}
For Rytov approximation, the update formula is derived based on the relation given in Eqs.\,\ref{eq5} and \ref{eq7}
\begin{equation}
\bar U_{s1n}^i({{\bf{x}}_T}) = \left\{ \begin{array}{l}
\ln \left( {\sqrt {I_c^i\left( {{\bf{x}}_T^{}} \right)} \frac{{\exp \left( {U_{s1n}^i({{\bf{x}}_T})} \right){\kern 1pt} }}{{\left| {\exp \left( {U_{s1n}^i({{\bf{x}}_T})} \right)} \right|}}} \right){\kern 1pt} {\kern 1pt} {\kern 1pt} {\kern 1pt} {\kern 1pt} {\kern 1pt} {\kern 1pt} {\kern 1pt} {\kern 1pt} {\kern 1pt} {\kern 1pt} {\kern 1pt} {\kern 1pt} {\kern 1pt} {\kern 1pt} {\kern 1pt} {\kern 1pt} {\kern 1pt} {\kern 1pt} {\kern 1pt} {\kern 1pt} {\kern 1pt} {\kern 1pt} {\kern 1pt} {\kern 1pt} {\kern 1pt} {\kern 1pt} {\kern 1pt} {\kern 1pt} {\kern 1pt} {\kern 1pt} {\kern 1pt} {\kern 1pt} {\kern 1pt} {\kern 1pt} {\kern 1pt} {\kern 1pt} {\kern 1pt} {\kern 1pt} {\kern 1pt} {\kern 1pt} {\kern 1pt} {\kern 1pt} {\kern 1pt} {\kern 1pt} {\kern 1pt} {\kern 1pt} {\kern 1pt} {\kern 1pt} {\kern 1pt} {\kern 1pt} {\kern 1pt} {\kern 1pt} {\kern 1pt} {\kern 1pt} {\kern 1pt} {\kern 1pt} {\kern 1pt} {\kern 1pt} {\kern 1pt} {\kern 1pt} {\kern 1pt} {\kern 1pt} {\kern 1pt} {\kern 1pt} {\kern 1pt} {\kern 1pt} {\kern 1pt} {\kern 1pt} {\kern 1pt} {\kern 1pt} {\kern 1pt} {\kern 1pt} {\kern 1pt} {\kern 1pt} {\kern 1pt} {\kern 1pt} {\kern 1pt} {\kern 1pt} {\kern 1pt} {\kern 1pt} {\kern 1pt} {\kern 1pt} {\kern 1pt} {\kern 1pt} {\kern 1pt} {\kern 1pt} {\kern 1pt} {\kern 1pt} {\kern 1pt} {\kern 1pt} {\kern 1pt} {\kern 1pt} {\kern 1pt} {\kern 1pt} {\kern 1pt} {\kern 1pt} {\kern 1pt} Bright{\kern 1pt} field\\
\ln \left( {\sqrt {I_c^i\left( {{\bf{x}}_T^{}} \right)} \frac{{\exp \left( {U_{s1n}^i({{\bf{x}}_T})} \right) - {\kern 1pt} {\kern 1pt} {\kern 1pt} {\kern 1pt} 1{\kern 1pt} }}{{\left| {\exp \left( {U_{s1n}^i({{\bf{x}}_T})} \right) - {\kern 1pt} {\kern 1pt} 1} \right|}} + 1} \right){\kern 1pt} {\kern 1pt} {\kern 1pt} {\kern 1pt} {\kern 1pt} {\kern 1pt} {\kern 1pt} {\kern 1pt} {\kern 1pt} {\kern 1pt} {\kern 1pt} {\kern 1pt} {\kern 1pt} {\kern 1pt} {\kern 1pt} {\kern 1pt} {\kern 1pt} {\kern 1pt} {\kern 1pt} {\kern 1pt} {\kern 1pt} {\kern 1pt} {\kern 1pt} {\kern 1pt} {\kern 1pt} {\kern 1pt} {\kern 1pt} {\kern 1pt} {\kern 1pt} {\kern 1pt} {\kern 1pt} {\kern 1pt} {\kern 1pt} {\kern 1pt} {\kern 1pt} {\kern 1pt} {\kern 1pt} {\kern 1pt} {\kern 1pt} {\kern 1pt} {\kern 1pt} {\kern 1pt} {\kern 1pt} {\kern 1pt} {\kern 1pt} {\kern 1pt} {\kern 1pt} {\kern 1pt} {\kern 1pt} {\kern 1pt} {\kern 1pt} {\kern 1pt} {\kern 1pt} {\kern 1pt} {\kern 1pt} {\kern 1pt} {\kern 1pt} {\kern 1pt} {\kern 1pt} {\kern 1pt} {\kern 1pt} {\kern 1pt} {\kern 1pt} {\kern 1pt}  Dark{\kern 1pt} field
\end{array} \right.
\label{eq9}
\end{equation}
     Unless otherwise noted, the Rytov approximation is employed for the FPDT reconstruction due to its higher RI reconstruction accuracy (see \textbf{Supplementary Section E} for a detailed comparison).
 \item Fourier transform this amplitude-constrained estimate, ${\bar U_{s1n}^i({{\bf{x}}_T})}$, to the frequency domain. The resultant spectrum is further shifted back to its originally position with a translation vector ${ - {\bf{k}}_{inT}^i =  - (k_{inx{\kern 1pt} }^i,k_{iny{\kern 1pt} }^i)}$. This forms an update of the first-order scattered field, ${\bar U_{s1}^i({{\bf{k}}_T})}$. Based on the values of ${\bar {\hat f}_{}^i\left( {{\bf{k}}_T^{}} \right) = -\frac{{{k_z}}}{{\pi j}}\bar {\hat U}_{s1}^i({{\bf{k}}_T})}$, we locally update the corresponding sub-region of the 3D spectrum ${\hat f\left( {\bf{k}} \right)}$  enclosed in the 3D generalized aperture, which is just the spherical cap extracted in Step 2. This completes one sub-iteration of the FPDT algorithm.
\end{enumerate}
Next, we move to the next illumination angle, which corresponds to a new spectral region. Steps 2-5 are then repeated until all $N$ images are scanned, and the whole iteration scheme is repeated over $M$ cycles to achieve a self-consistent solution. To accelerate and stabilize the convergence of the algorithm, we use adaptive relaxation strategy to gradually diminish the updating weight as the iteration accumulates\cite{39}. At the end of this iterative recovery process, the converged solution in Fourier space will typically cover a significantly extended spectral support. With our current configuration, we can acquire an image within illumination angles of up to ±64.2\degree (see \textbf{FPDT platform and characterization}). As a result, the entire region of frequency space still cannot be filled. Thus, in the final step of the reconstruction algorithm, we use an iterative constraint algorithm\cite{24,36} to computationally fill this missing information, which mitigates the elongation of the reconstructed shape along the optical axis and generates a more accurate estimation of the 3D RI (see \textbf{Methods}).
\begin{figure}
  \centering
  \includegraphics[width={1.0\linewidth}]{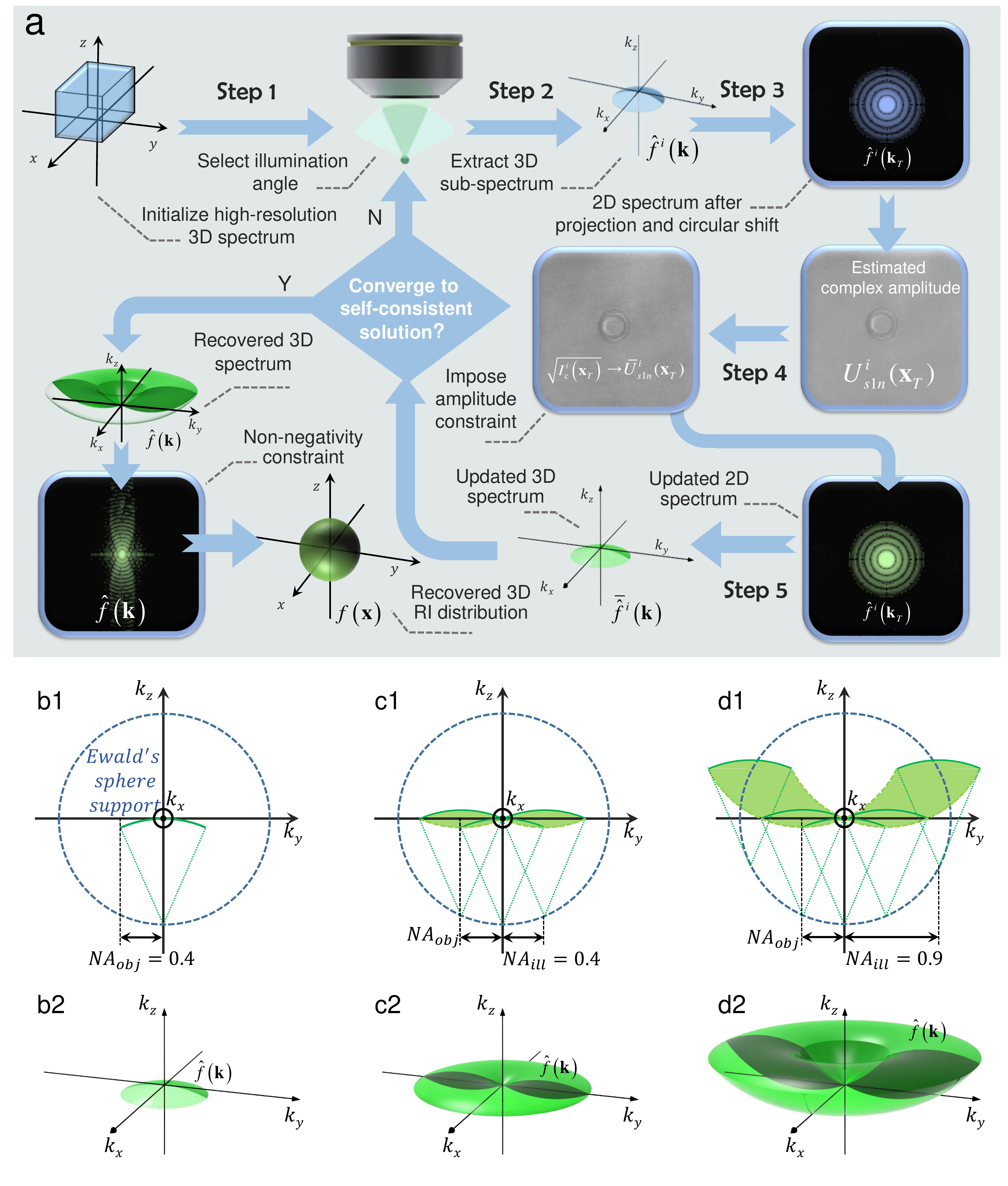}
  \caption{FPDT reconstruction algorithm and frequency coverage for different illumination configurations.
  a, Flow chart of the entire FPDT reconstruction algorithm.
  b, Accessible object frequencies when ${{NA}{_{obj}} = 0.4}$ and ${{NA}{_{ill}} = 0}$  (coherent illumination).
  c, Accessible object frequencies when ${{NA}{_{obj}} = 0.4}$ and ${{NA}{_{ill}} = 0.4}$ (matched bright-field illumination).
  d, Accessible object frequencies when ${{NA}{_{obj}} = 0.4}$ and ${{NA}{_{ill}} = 0.9}$ (both bright-field and dark-field illuminations are used).
  b1-d1, 2D spectral supports in the $k_y$-$k_z$ planes for different illumination configurations.
  b2-d2, corresponding 3D spectral supports for different illumination configurations.}
  \label{fig3}
\end{figure}

\section*{Frequency coverage and resolution analysis}
In FPDT, the support for the accessible object frequencies can be used to assess both lateral and axial resolution limit theoretically, as illustrated in Fig.\,3b,c for different illumination configurations. Without loss of generality, here we consider a microscope in air with a rotationally symmetric aperture, which allows us to limit our attenuation to only one cross-section (y-z slice) of the 3D CTF.

For a coherent microscope with a fixed orthogonal plane wave illumination, the accessible object frequency just corresponds to the support of the generalized aperture, with a half-side lateral frequency extent of ${{{{NA}{_{obj}}} \mathord{\left/
 {\vphantom {{{NA}{_{obj}}} \lambda }} \right.
 \kern-\nulldelimiterspace} \lambda }}$  and half-side axial extent of ${{{\left( {1 - \sqrt {1 - {NA}_{obj}^2} } \right)} \mathord{\left/
 {\vphantom {{\left( {1 - \sqrt {1 - {NA}_{obj}^2} } \right)} \lambda }} \right.
 \kern-\nulldelimiterspace} \lambda }}$, as illustrated in Fig.\,\ref{fig3}b. The 3D CTF is only a fraction of the Eward sphere with no axial support, resulting in a very limited discrimination along the optical axis\cite{40,41}. ODT techniques circumvent this issue by sequentially illuminating a 3D object from various directions to enlarge the accessible 3D Fourier domain. In a conventional ODT system, the maximum illumination angle allowed is limited to the maximum collection NA, forming a torus-shaped structure in 3D frequency space\cite{21,32} (Fig.\,\ref{fig3}c). The half-side lateral and axial extensions of the frequency supports are given by
\begin{equation}
\label{eq10}
\Delta {u_{x,y}} = \frac{{{NA}{_{obj}}}}{\lambda }{\kern 1pt} {\kern 1pt}, {\kern 1pt} {\kern 1pt} {\kern 1pt} {\kern 1pt} {\kern 1pt} {\kern 1pt} \Delta {u_z} = \frac{{1 - \sqrt {1 - {NA}_{obj}^2} }}{\lambda }
\end{equation}
Note that the frequency support of a conventional ODT system is the same as that obtained in a conventional incoherent illuminated microscope, with a doubled lateral resolution compared to the coherent diffraction limit\cite{21,40,41}. Furthermore, the sequential synthetic aperture procedure also allows the transmission of all spatial frequencies within the support without any attenuation (the transfer function of ODT is always 1 within its support), whereas the optical transfer function (OTF) of a conventional incoherent microscope strongly dims high spatial frequencies\cite{42}. However, the limited angular coverage of the incident beam leads to a very restricted frequency support, which yields a relatively low axial resolution especially when a low NA objective is used. As the case shown in Fig.\,\ref{fig3}c, when ${{NA}{_{obj}} = 0.4}$  and ${\lambda = 507 nm}$, the depth resolution is ${{1 \mathord{\left/
 {\vphantom {1 {\Delta {u_z}}}} \right.
 \kern-\nulldelimiterspace} {\Delta {u_z}}} \approx 6.07 \mu m}$, which is almost one order of magnitude lower than the lateral resolution of ${{1 \mathord{\left/
 {\vphantom {1 {\Delta {u_{x,y}}}}} \right.
 \kern-\nulldelimiterspace} {\Delta {u_{x,y}}}} \approx 634 nm}$.

 If the synthetic aperture process can further incorporate high-angle dark-field illuminations that can be much greater than that allowed by the NA of the objective, the accessible object frequency will be improved significantly, as shown in Fig.\,\ref{fig3}c. By processing these dark-field measurements and inverting the data into an image, we can in principle obtain a synthetic transfer function that can be many times larger than the usable lens transfer function. As illustrated in Fig.\,\ref{fig3}d, the half-side lateral and axial extensions of the resultant frequency supports are
 \begin{equation}
 \label{eq11}
   \Delta {u_{x,y}} = \frac{{{NA}{_{obj}}{\rm{ + }}{NA}{_{ill}}}}{\lambda }{\kern 1pt} {\kern 1pt}, {\kern 1pt} {\kern 1pt} {\kern 1pt} {\kern 1pt} {\kern 1pt} {\kern 1pt} \Delta {u_z} = \frac{{1 - \sqrt {1 - {NA}_{ill}^2} }}{\lambda }
 \end{equation}
 It can be calculated that by employing high-angle illumination with ${{NA}{_{ill}} = 0.9}$  which is much larger than ${{NA}{_{obj}} = 0.4}$, the lateral and axial resolution can be increased to 390$nm$ and 899$nm$ respectively, suggesting a remarkable resolution enhancement in both lateral and axial directions. In addition, the accessible frequency support (the volume of the non-zero 3D spectral region) is expanded to 9.2 times its original volume, resulting in a significant increase in the SBP. However, the dark-field scattered light is generally very weak and creates severe speckle noise in the case of coherent illumination, preventing reliable interferometric measurement of the phase component. This challenge can be effectively bypassed by our FPDT method since it only utilizes a standard microscope to detect image intensities for different angled illuminations without requiring direct phase measurement.

\section*{FPDT platform and characterization}
\begin{figure}
  \centering
  \includegraphics[width={1.0\linewidth}]{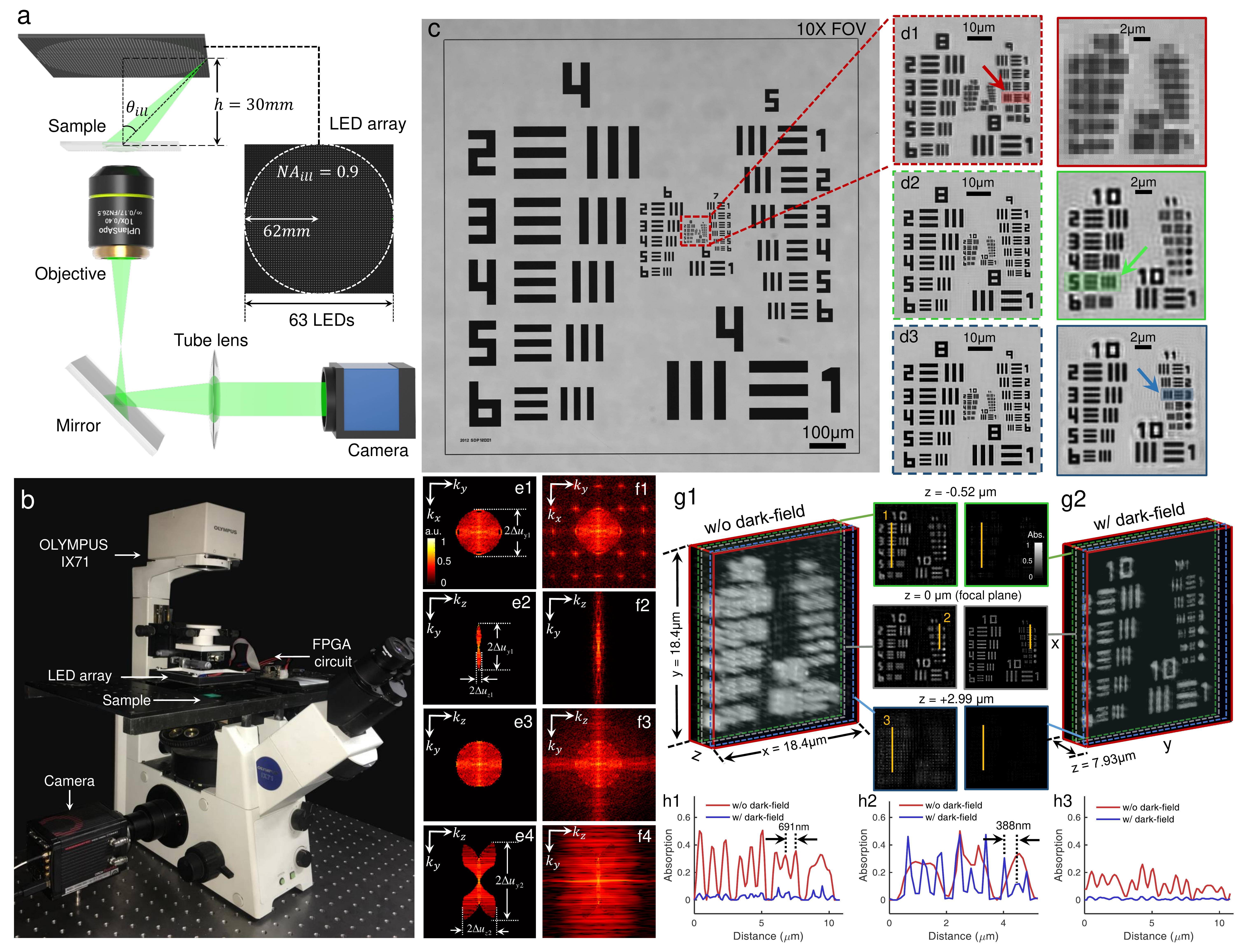}
  \caption{Optical setup of FPDT platform and its characterization.
  a, Schematic diagram of the illumination and imaging system of the FPDT platform. The light source of an off-the-shelf inverted microscope (IX71, Olympus, Japan) is replaced by an LED array.
  b, A photograph of the FPDT platform.
  c, Bright-field low-resolution raw image acquired with the full-FOV of a 10$\times$ objective.
  d1, Magnified raw image corresponding to the red-boxed central part of the USAF target.
  d2, High-resolution recovery result using ordinary FPM method without using dark-field images.
  d3, High-resolution recovery result of the same sub-region using ordinary FPM method with all of the raw images, including dark-field images.
  e1-e2, Lateral and axial sections of the recovered high-resolution 3D absorption spectrum without using dark-field images.
  e3-e4, Lateral and axial sections section of the recovered high-resolution 3D absorption spectrum using all of the raw images.
  f1-f2, Lateral and axial sections of the recovered high-resolution 3D absorption spectrum without using dark-field images after the iterative constraint algorithm.
  f3-f4, Lateral and axial sections section of the recovered high-resolution 3D absorption spectrum using all of the raw images after the iterative constraint algorithm.
  g1-g2, Volume-rendered image of the USAF target reconstructed by FPDT with and without using dark-field images. Three pairs of absorption slices taken at depths of -0.52, 0, and 2.99$\mu m$.
  h1-h2, The corresponding line profiles are shown to compare the imaging resolution (indicated by the yellow lines in g1-g2, respectively).}
  \label{fig4}
\end{figure}

As depicted in Fig.\,\ref{fig4}a,b, the FPDT platform consists of two major components: a programmable LED array and a conventional microscopic imaging system (see \textbf{Methods}). The surface-mounted LED array (2 $mm$ spacing) is placed 30 $mm$ above the object plane, where 3001 LED elements from a 63$\times$ 63 LED array are used to create quasi-monochromatic (central wavelength of 507$nm$), spatially coherent quasi-plane waves from different angles with a maximum illumination NA of 0.9. The image is captured by an inverted microscopic system (IX71, Olympus, Japan) with a 10$\times$ 0.4NA objective (UPLSAPO 10$\times$, Olympus, Japan) and a scientific complementary metal-oxide-semiconductor (CMOS) camera (Hamamatsu ORCA-Flash 4.0 C13440, pixel resolution 2048$\times$2048 , pixel pitch 6.5$\mu m$ ).

We demonstrate the imaging performance of our FPDT platform experimentally by measuring a 1951 USAF resolution target (Ready Optics Company, USA). First, the resolution target was horizontally placed on the sample stage of the microscope. Figure \ref{fig4}c shows the raw full-FOV image of the object under axial illumination from the central LED, and red-boxed region (64 $\times$ 64 pixels), which contains the smallest groups of features (Groups 10 and 11), is extracted and shown in Fig.\,\ref{fig4}d. Under coherent illumination, the final resolution is limited by the pixel size of the camera (650$nm$ at the object plane), instead of the coherent diffraction limit (1.27$\mu m$, Group 9, Element 5), creating significant aliasing effects (pixelization). Since the flat USAF target can be regarded as a thin 2D object, we first used the conventional FPM algorithm to recover the 2D super-resolved images of this target and calibrated the spatial position of each LED element numerically\cite{43}. Figure \ref{fig4}d2,3 presents the images recovered by FPM without or with dark-field raw images, revealing that the highest resolution achieved is enhanced from 616$nm$ (Group 10, Element 5) to 388$nm$ (Group 11, Element 3), which agrees well with the theoretical resolution limit of 634$nm$ (${{NA}{_{syn}} = 0.8}$ ) and 390$nm$ (${{NA}{_{syn}} = 1.3}$), respectively.

Next, the same data set was used to implement 3D tomography based on FPDT. When only bright-field images are used for the reconstruction, the theoretical lateral and axial resolution limits can be determined as 634$nm$ and 6.07$\mu m$, respectively, according to the lateral and axial sections of the recovered high-resolution 3D absorption spectrum shown in Fig. 4e1,2. Since the recovered information occupies only a very limited portion of the whole spectrum, the missing information can hardly be compensated by the iterative algorithm, leading to significant grid artefacts (Fig.\,\ref{fig4}f1,2). When all the raw images, including the dark-field images, were used for the 3D tomographic reconstruction, the accessible frequency support was significantly expanded, as illustrated in Fig.\,\ref{fig4}e3,4. In addition, the missing cone problem in the 3D Fourier spectrum was significantly alleviated by the iterative algorithm, as shown in Fig.\,\ref{fig4}f3,4. Figure \ref{fig4}g and \textbf{Supplementary Movie 1} show the volume-rendered absorption images and absorption sections for different axial planes. These results together with the corresponding line profiles shown in Fig.\,\ref{fig4}h1,3 confirm that the dark field measurements improve the lateral resolution from 691$nm$ (Group 10, Element 4) to 388 $nm$ (Group 11, Element 3). In addition, the axial elongation problem is significantly alleviated: the object information is completely indistinguishable when the defocus distance is 0.52$\mu m$, which is almost one-six of that without using the dark field images (2.99$\mu m$), confirming a depth resolution within the 1$\mu m$ range (in accordance with the theoretical prediction). We also evaluated the depth sectioning capability of the FPDT technique by tilting the USAF target with a small angle ($\sim$10\degree). The results shown in \textbf{Supplementary Movie 2} and \textbf{Supplementary Section C} demonstrate that the out-of-focus features can be successfully rejected and only the in-focus information appears in the reconstructed sections. In \textbf{Supplementary Section D} and \textbf{Supplementary Figure 3}, we further determine the depth of focus (DOF) of the proposed FPDT platform to be $\sim$20$\mu m$ without significantly compromising the lateral resolution, which is 5-fold longer than the 4$\mu m$ natural DOF associated with the 10$\times$ 0.4NA objective used in the experiment, and $\sim$40-fold longer than that of a conventional microscope objective with the same NA of 1.3 .

\section*{3D tomographic imaging of Pandorina}
\begin{figure}
  \centering
  \includegraphics[width={1.0\linewidth}]{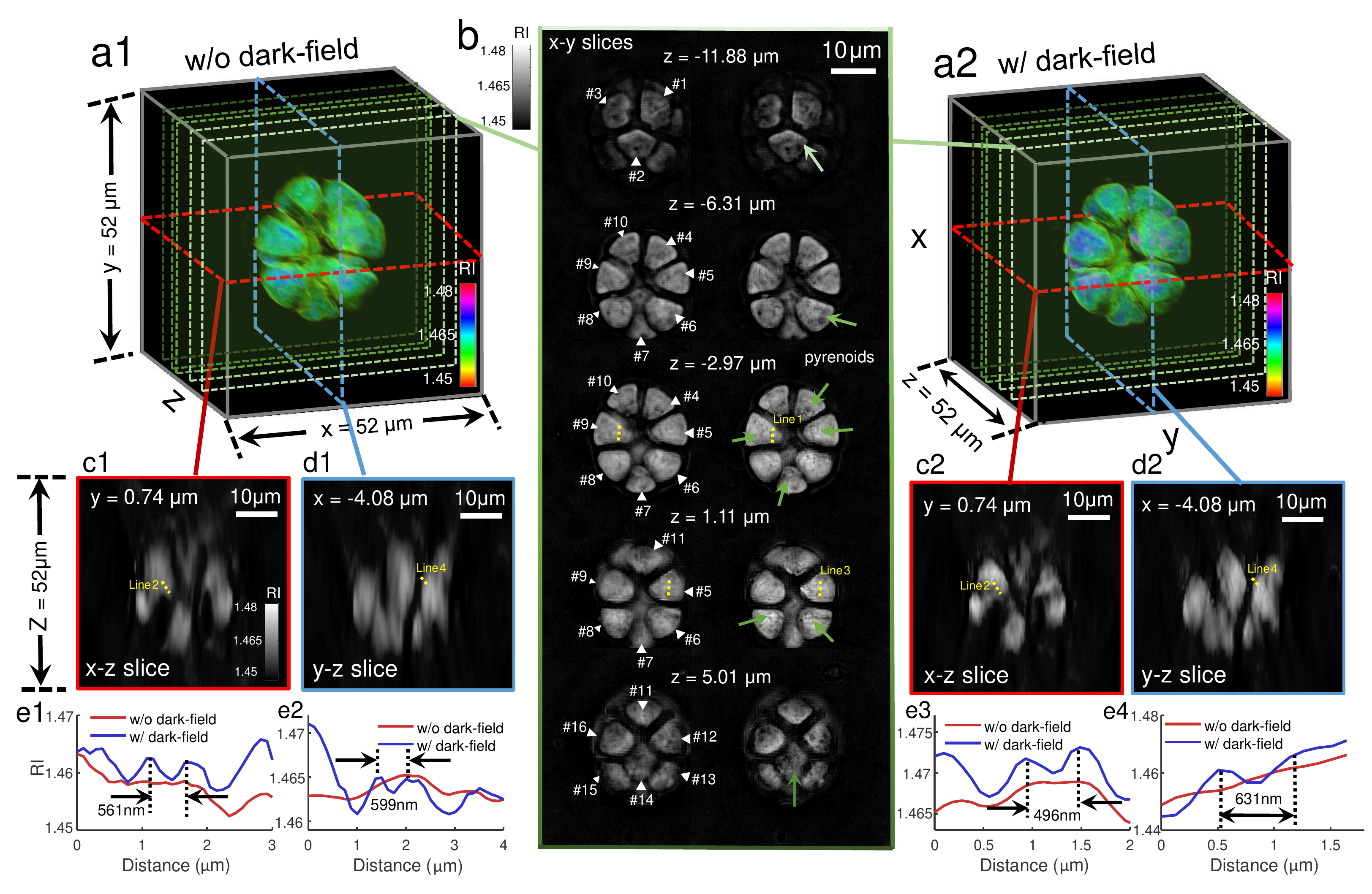}
  \caption{3D quantitative RI reconstruction of a \textsl{P. morum} using FPDT.
  a1-a2, 3D rendering results of the RI of the \textsl{P. morum} using the FPDT algorithm (without and with using dark-field imaging) in x-y, y-z and x-z directions.
  b, Five 2D slices of the RI distribution at different depths recovered using FPDT without and with dark-field imaging.
  c1-c2, d1-d2, Two RI slices along the x-z and y-z directions recovered using FPDT without and with using dark-field imaging.
  e1-e4, The corresponding line profiles are shown to compare the imaging resolution (indicated by the yellow lines in b-d, respectively).}
  \label{fig5}
\end{figure}

In this section, the proposed FPDT technique is applied to a bleached paraffin section of Pandorina morum algae (\textsl{P. morum}). Figure \ref{fig5} displays the 3D rendering results of a 16-celled \textsl{P. morum} using Rytov approximation based FPDT algorithm (with and without using dark-field imaging) in x-y, y-z and x-z directions. In addition, the recovered through-slice RI stacks of the \textsl{P. morum} and the corresponding 3D rendered RI images are animated in \textbf{Supplementary Movie 3}. In these results, the inner architecture of the algae is clearly visualized, and we can see clearly how different cells are held together with respect to each other in 3D space (each individual cell is indicated by white arrows in Fig.\,\ref{fig5}b). When the dark-field images are used in the FPDT technique, more subcellular details inside the \textsl{P. morum}, e.g. the periphery of pyrenoids, can be revealed (indicated by the green arrows in Fig.\,\ref{fig5}b), as illustrated in the x-y slices of RI distribution images at 5 different depths (Fig.\,\ref{fig5}b). In addition, the vertical cellular structure became more clear and compact, as can be observed in the x-z (Fig. 5c1,c2) and y-z slices (Fig.\,\ref{fig5}d1,d2). The line profiles shown in Fig.\,\ref{fig5}e1-e4 confirm that more fine structures within the algae can be resolved by incorporating the dark-field measurements. In \textbf{Supplementary Section F} and \textbf{Supplementary Figure 6}, we further demonstrate the tomographic imaging results of another algae, \textsl{C. hibernicus} diatom. The recovered through-slice RI stacks and corresponding 3D rendered RI images are animated in \textbf{Supplementary Movie 5}. Once again, our FPDT clearly reveals its fine intracelluar features such as the spines and pores covering the valve face when high-NA dark-field measurements are incorporated.

\section*{Wide-field tomographic imaging of unstained HeLa cells}
Finally, to demonstrate the performance of the FPDT approach for wide-FOV and high-resolution 3D imaging in life sciences, we imaged a large population of adherent HeLa cells. To preserve the cell morphology, HeLa cells were fixed in phosphate-buffered saline (PBS) buffer (${n_m} = 1.34$) on a microscope slide. Figure\,\ref{fig5}a shows the full-FOV RI slice of the fixed HeLa cells slide at z = 2.93$\mu m$ reconstructed using the FPDT algorithm (with dark-field measurements). To illustrate the subcellular structures inside individual HeLa cells, the RI slices of three zoomed areas (62.4 $\times$ 62.4 $\mu m$) enclosing three typical cells within the entire FOV are selected and shown in Fig. \,\ref{fig5}b1-b3. Their corresponding rendered x-y, y-z, x-z projections are displayed in Fig.\,\ref{fig5}c1-c3, respectively. \textbf{Supplementary Movie 4} further illustrates 360\degree 3D renderings of these three cells and other depth sections of the whole sample. In all these results, the optically dense nucleus, cytoskeletal fibres, and cytoplasmic organelles are shown with high contrast, and the appearance of distinct details at different sections can be clearly observed, suggesting excellent sectioning ability that is otherwise unattainable with conventional QPI approaches. In \textbf{Supplementary Section G} and \textbf{Supplementary Figure 7}, we further demonstrate the tomographic imaging results of an unstained blood smear. The recovered full-FOV through-slice RI stacks and 3D renderings of the RI distribution of three selected sub-regions (200 $\times$ 200 $\mu m^2$) are animated in \textbf{Supplementary Movie 6}. The full-FOV 3D reconstruction contains $\sim$17.2 billion voxels of quantitative 3D RI data with $\sim$20,000 blood cells over a 1.77$mm^2$ FOV, revealing the potential of our FPDT approach and platform for high-content quantitative analysis of a large population of cells, which is of utmost importance for many applications, such as cancer screening, stem cell research, and drug development.
\begin{figure}
  \centering
  \includegraphics[width={1.0\linewidth}]{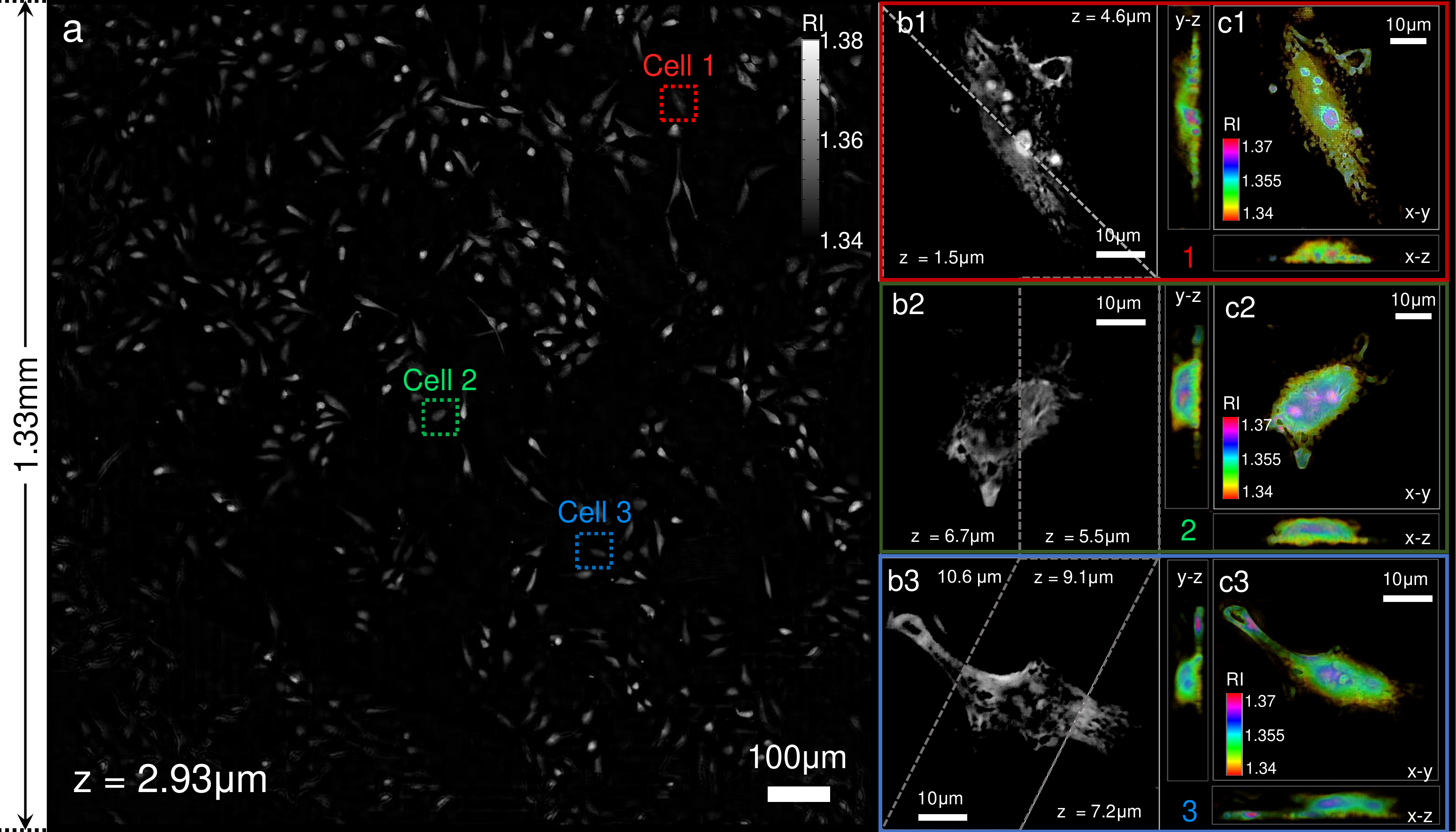}
  \caption{Wide-FOV and high-resolution 3D imaging of a large population of HeLa cells using FPDT.
  a, Full-FOV RI slice of the fixed HeLa cells slide at z = 2.93$\mu m$ reconstructed using the FPDT algorithm.
  b1-b3, RI slices of three zoomed areas enclosing three typical cells within the entire FOV at different depths.
  c1-c3, Rendered x-y, y-z, x-z RI projections corresponding to the selected 3 HeLa cells.}
  \label{fig6}
\end{figure}

\section*{Conclusion and discussion}
In this study, we have demonstrated FPDT, which is based on intensity-only measurements captured by an off-the-shelf microscope with a simple LED array source add-on. By synthesizing a set of variably illuminated, low-resolution intensity images acquired with a low-NA objective in 3D Fourier space, a wide-field, high-resolution depth-resolved complex RI image across large volumes can be reconstructed. There are two distinct features of FPDT: First, without resorting to holographic phase measurement, FPDT achieves phase retrieval, synthetic aperture, and tomographic reconstruction from low-NA intensity-only measurements simultaneously. Second, FPDT incorporates high-angle dark-field illuminations (up to 0.9NA) to significantly expand the accessible object frequency. As a result, the FPDT platform offers the highest-throughput ODT results with 390$nm$ lateral resolution together with an axial resolution of 899$nm$ across a 10$\times$ FOV of 1.77$mm^2$  and a DOF of $\sim$20$\mu m$, creating an effective voxel size of 0.0043$\mu m^3$ across a sample volume of 0.074$mm^3$. The tomographic imaging performance of FPDT has been quantified by imaging a USAF resolution target and demonstrated by quantitative measurements of the complex RI through several types of thick specimens. The experimental results suggest that the proposed FPDT is an effective and promising tool for large-scale high throughput 3D bio-imaging in a label-free fashion.

As a recording and post-processing technique, FPDT has not yet been fully optimized in terms of processing speed. The large data requirement ($\sim$12.6 gigapixels per whole FOV dataset) creates a severe burden on both storage and processing. We have implemented the FPDT reconstruction in MATLAB. The time required to reconstruct the entire tomogram is $\sim$23 hours on a workstation. The processing speed may be significantly improved to several minutes by using graphics processing units (GPUs) with larger graphics memory, since each smaller sub-tomogram can be independently and parallelly processed. Furthermore, the FPDT method requires much higher redundancy than the FPM method\cite{44} since it requires sufficient overlap of the generalized aperture in 3D Fourier spectrum to guarantee fast and stable convergence. Therefore, the overall size of the data set is significantly larger than that of the FPM. In addition, in our current system, the distribution of LED elements in the array suggests that the high-frequency dark-field region of the 3D object spectrum has a larger overlapping ratio than that of the central bright-field region. Therefore, it is expected that reconstructions can still be successful with much fewer intensity measurements by properly redesigning the LED illuminator\cite{45,46} or carefully selecting the active LEDs in the array\cite{44}. Finally, it would be worth exploring the multiplexed illumination strategies\cite{47,48} and the recently emerged deep learning reconstruction approaches\cite{49,50} to further reduce the data size and speed up the acquisition procedure. By merging these approaches, it may be possible to increase the image acquisition speed from 91 seconds per 3D frame to less than 10 seconds per 3D frame without incurring data management issues, enabling the recording of 3D videos of subcellular dynamical phenomena over a wide FOV for time lapse \emph{in vitro} microscopy applications.

\begin{methods}
\subsection{Iterative constraint algorithm}
To minimize the artefact introduced by the missing angle space, we applied an iterative constraint algorithm based on prior knowledge that the RI of the object is higher than that of the surrounding medium, and the imaginary part of the complex RI should always be no less than zero (no negative absorption). We first take an inverse 3D Fourier transform of the originally retrieved spectral data with zero values for the missing space (Fig.\,\ref{fig3}a). In the spatial domain, there are pixels whose index values are smaller than RI of the medium, we force these to be the same as the RI of the medium\cite{24,36}. In addition, for pixels whose absorption coefficients are smaller than zero, we also set these to zero to prevent negative absorption. Then we take a 3D Fourier transform and obtain the updated spectral data in which the missing space is no longer zero. In this way, we obtain an approximate solution for the missing angles. At the same time, the data within the original frequency support are replaced with the initial retrieved data. The rest of the data remains unchanged. This constitutes a single iteration of the iterative constraint algorithm. We iterate this procedure until the reconstructed object function converges. Usually, up to 50 iterations are required to guarantee the convergence. Finally, the negative bias in the complex RI is removed and the reconstructed object function becomes more accurate.

\subsection{Experimental setup}
The FPDT platform is built based on a conventional microscopic imaging system with the light source replaced by a programmable LED array. The central wavelength of the LED illumination is 507$nm$, and the spectral bandwidth is $\sim$20$nm$. During the data collection procedure, 3001 LED elements from the 63$\times$63 LED array are lighted up sequentially, creating quasi-monochromatic, spatially coherent quasi-plane waves from different angles with a maximum illumination NA of 0.9. The LED array is driven by a self-developed circuit with a field programmable gate array (FPGA) unit (EP4CE10E22C8N, ALTERA, US) to provide the logical control. The microscopic imaging system consists a commercial bright-field microscope (IX71, Olympus, Japan) including a 10$\times$ 0.4NA objective (UPLSAPO 10$\times$, Olympus, Japan), and a scientific CMOS camera (Hamamatsu ORCA-Flash 4.0 C13440, pixel resolution 2048$\times$2048,  pixel pitch 6.5$\mu m$). The camera is synchronized with the LED array by the same controller via two coaxial cables that provide the trigger and monitor the exposure status. We experimentally measure the system frame rate to be $\sim$33Hz for capturing full-frame (2048$\times$2048) 16-bit images. Thus, all those 3001 images are captured within 91 seconds. The data is transferred to the computer via a CameraLink interface.

\subsection{Practical implementation of FPDT}
The FPDT reconstruction algorithm is implemented based on MATLAB software (MATLAB R2016a) with a computer workstation (Intel Core i7-7800X, 3.50 GHz central processing unit, 64 GB random-access memory, NVIDIA GeForce GTX 1080Ti 8GB graphics card). In this work, because we focused on the proof-of-concept, the image reconstruction speed was not optimized. The time required for all computation processes (including 20 FPDT iterations and 50 iterations of the non-negativity constraint) of intensity stacks within a small segment (64 $\times$ 64 $\times$ 3001) from the whole FOV  (2048$\times$2048$\times$3001)  is  $\sim$51 seconds. Thus, for the entire FOV of the digital camera, we divided each full-FOV raw image (2048 $\times$ 2048) into 40 $\times$ 40 sub-regions (64 $\times$ 64 pixels each) with a 13-pixel overlap on each side of neighboring sub-regions. After obtaining the high-resolution tomogram of each sub-region, a full-FOV high-resolution 3D tomography result (8192$\times$8192$\times$256) is created by using an alpha-blending stitching method. The total processing time for the full-FOV is $\sim$22.7 hours, which could in fact be significantly reduced by investigating the use of GPU acceleration. More detailed discussions about the computational efficiency of the FPDT reconstruction algorithm can be found in \textbf{Supplementary Section B} and \textbf{Supplementary Figure 1}.
\end{methods}



\begin{addendum}
 \item The authors thank X. Wang for preparing the cell samples, X. Pan for developing the hardware controller, J. Ding for help with the schematic drawing, and R. Zhang for editing the manuscript. This work was sponsored in part by National Natural Science Foundation of China (61722506, 11574152), Outstanding Youth Foundation of Jiangsu Province (BK20170034), and The Key Research and Development Program of Jiangsu Province (BE2017162).
 \item[Author contributions] C. Z. and J. S. contributed equally to this work. C. Z. initiated the project and developed the theory and method. C.Z. wrote code for the experiment and simulations. C.Z. and J.S. designed the experiments. J.S. and J.L built the experimental platform. J.S. and J.L prepared and performed the experiments. C. Z. and J. S. analysed the data. J. S. prepared the supplementary movies. A. A. and Q. C provided research advice. C. Z, A. A. and Q. C provided overall supervision. C.Z. and J.S. wrote the manuscript with contributions from all authors.
 \item[Competing Interests] J. S., J. L., and A. A. declare no competing financial interests. C. Z and Q. C are named inventors on several related patent applications. C. Z and Q. C also have competing financial interests in Nanjing Jiangnan Novel Optics Co., Ltd. and Suzhou Flyingman Precision Instrument Co., Ltd. China, which, however, did not support this work.
\end{addendum}


\end{document}